\documentclass[10pt]{article}
\usepackage{amsmath,amsthm,amssymb}
\usepackage[cp1251]{inputenc}
\usepackage[russian]{babel}
\usepackage{graphicx}
\usepackage[small]{caption2}
\usepackage{indentfirst}
\usepackage{amsmath}
\usepackage{afterpage}

\usepackage{epsfig}
\usepackage{epstopdf}

\usepackage[usenames]{color}

\usepackage{amssymb,amsmath}
\usepackage{amsfonts}
\usepackage{epsfig}
\usepackage{ulem}

\usepackage{cite}
\def\gee{ \, \lower 1mm\hbox{$\,{\buildrel > \over{\scriptstyle\scriptstyle\sim} }\displaystyle \,$}}
\def\lee{ \, \lower 1mm\hbox{$\,{\buildrel < \over{\scriptstyle\scriptstyle\sim} }\displaystyle \,$}}
\def\|{\partial}

\def\varkappa {{\scriptstyle\partial}\! e}

\def\namepoint{$\!\!\!\!\!\!\!{\textbf{.}}\,$}

\let\b=\baselineskip
\begin{document}
\renewcommand{\captionlabeldelim}{.}
\headheight 1.50true cm \headsep  0.7true cm \righthyphenmin=2

\parindent=10.5mm

\marginparwidth=20true mm

 \noindent{\it
D'yakonova T.A., Khrapov S.S., Khoperskov A.V. The problem of boundary conditions for the shallow water equations, Vestnik Udmurtskogo Universiteta: Matematika, Mekhanika, Komp'yuternye Nauki, 2016, vol.26, no.3, pp.401-417}

\ 

\centerline{\textbf{The problem of boundary conditions for the shallow water equations (Russian)}}

\bigskip
\centerline{T.\,A.~Dyakonova, S.\,S.~Khrapov, A.\,V.~Khoperskov $^1$}

\bigskip
$^1$ --- Volgograd State University

\bigskip
{\textbf{Abstract}}

The problem of choice of boundary conditions are discussed for the case of numerical integration of the shallow water equations on a substantially irregular relief. In modeling of unsteady surface water flows has a dynamic boundary partitioning liquid and dry bottom. The situation is complicated by the emergence of sub- and supercritical flow regimes for the problems of seasonal floodplain flooding, flash floods, tsunami landfalls. Analysis of the use of various methods of setting conditions for the physical quantities of liquid when the settlement of the boundary shows the advantages of using the waterfall type conditions in the presence of strong inhomogeneities landforms. When there is a waterfall on the border of the computational domain and heterogeneity of the relief in the vicinity of the boundary portion may occur, which is formed by the region of critical flow with the formation of a hydraulic jump, which greatly weakens the effect of the waterfall on the flow pattern upstream.


\bigskip


 \noindent
Дьяконова Т.А., Храпов С.С., Хоперсков А.В. Проблема граничных условий для уравнений мелкой воды // Вестник Удмуртского университета. Математика. Механика. Компьютерные науки. 2016. Т. 26. № 3. С. 401-417.

{\textbf{Введение}}



Решение широкого круга задач динамики поверхностных вод для самых различных приложений основывается на модели мелкой воды (уравнениях Сен-Венана) \cite{Shokin-etal:2015} и ее модификациях \cite{Fedotova-Khakimzjanov:2008,Green-Naghdi:1976,Bautin-Derjabin:2012,Peliotskij:1996,Bautin-etal:2010,Prokofev-2000}, включая многослойные модели\cite{Danilova:2014}, либо с учетом вертикальной структуры на примитивных уравнениях \cite{khakimzyanov2015simulation}, с учетом дисперсионных эффектов в различных приближениях \cite{Liapidevskii-1999,Zeytunyan:1995,Shokin-etal:2015}. Укажем на направления, связанные с моделированием русловых объектов --- водохранилищ \cite{Gusev-etal:2013} и речных систем \cite{Horritt-etal:2007foodplain,Bolgov-2014,Pisarev-etal:2013Udmurt,Caviedes-Voullieme-etal-2014}, крупных озер, дождевых потоков \cite{Costabile:2013}, динамики плотных мелкозернистых геофизических потоков \cite{Juez-etal-2014}, волн цунами \cite{Marchuk-Moshkalev:2014}.

 При наличии твердых границ мы имеем возможность корректного описания жидкости, как для Эйлеровых численных схем, так и в случае Лагранжевого подхода, применяя, например, метод модифицированных виртуальных граничных частиц \cite{Vacondio-etal:2012}. В случае озер или морей, если водный объект полностью находится в пределах расчетной сетки, проблема граничных условий по-сути отсутствует, если применять алгоритмы <<движения по сухому дну>> \cite{Bautin-etal:2011}. В случае протяженных систем (водохранилищ и рек) практически всегда имеются границы, через которые вода поступает в расчетную область и уходит из нее, что требует аккуратной постановки граничных условий \cite{Burguete-etal:2004Implicit-schemes,Burguete-etal:2006Numerical-boundary-conditions}. Наиболее сложным оказывается режим, при котором граница <<жидкость -- сухое дно>> достигает границы расчетной области.
 Такая ситуация имеется при численном исследовании динамики поверхностных стоков воды, возникающих при затоплении территории, движении мелкодисперсных потоков вещества \cite{Juez-etal-2014}, особенно на сложном горном рельефе \cite{Burguete-etal:2002extreme-rainfall}, например, в случае прорыва моренно-подпрудных приледниковых и ледниковых озерных систем \cite{Westoby-etal-2014}, из-за дождевых осадков \cite{Singh-etal-2014}.
Сходная проблема возникает при использовании уравнений мелкой воды для моделирования динамики атмосферы \cite{Skiba-1995}. При прохождении вещества через границу помимо численной устойчивости расчета важным является выполнение законов сохранения и построение неотражающих условий \cite{Ilgamov-Gilmanov-2003book}.

Хорошо известна проблема описания нестационарной линии уреза воды, связанная с необходимостью находить решение в области с подвижной границей (например, \cite{Bautin-etal:2010,Bautin-Derjabin:2012} для различных моделей мелкой воды \cite{Zokagoa-Soulaimani:2010}).
 В численных подходах при ее решении используются различные методы \cite{Liang-Borthwick-2009,Kopysov-2015}.
 В случае достаточно гладкой функции дна $b(x,y)$ можно успешно использовать регуляризацию \cite{Vater-etal-2015}.
 Использование модифицированного закона сохранения полного импульса с учетом образования локальных турбулентно-вихревых структур в поверхностном слое воды позволяет получить удовлетворительное согласие с экспериментальными данными \cite{Ostapenko-2007!dry-bed}.
 Полученные результаты аналитического исследования решений использованы для разработки новых аппроксимаций краевых условий на подвижной линии уреза. Однако, учитывая нерегулярный характер изменений реального дна $b(x,y)$ на различных масштабах, включая $L/H\sim 1$ ($1/L=|\nabla b| / b$, $H$~--- глубина жидкости), возникает необходимость строить более сложные алгоритмы, специально предназначенные для моделирования динамической границы <<жидкость -- сухое дно>> \cite{Khrapov-etal_csph-tvd:2011}.

Выделим ситуации, требующие аккуратной постановки граничных условий:
 1) жидкость свободно проходит через границу в докритическом режиме;
 2) вблизи свободной границы имеется участки сильно неоднородного рельефа;
 3) возникает поток жидкости внутрь расчетной области;
 4) происходят существенно нестационарные процессы вблизи границы.

  Целью данной работы является исследование влияния различных граничных условий в численной модели мелкой воды на структуру руслового потока.

\section{\label{t-p1}Основные уравнения и геометрия задачи}

Рассмотрим одномерную модель на основе системы уравнений мелкой воды с учетом силы придонного трения \cite{Zokagoa-Soulaimani:2010,Khrapov-etal:2013,Dyakonova-2014}:
\begin{gather}\label{eq-H}
\frac{\partial H}{\partial t} + \frac{\partial q}{\partial
x}=0,
\end{gather}
\begin{gather}\label{eq-q}
\frac{\partial q}{\partial t} + {\partial (q u)\over\partial x} = -gH {\partial \eta \over\partial x}
    - H \frac{\lambda}{2}\,u |u|,
\end{gather}
где $H$~--- толщина слоя жидкости, $q=Hu$~--- расход жидкости, $u=q/H$~--- скорость потока, $g=9.81$\,м/с$^2$~--- ускорение свободного падения, $\eta=H+b$~--- уровень свободной поверхности воды, $b(x)$~--- функция дна, $\lambda=2gn_M^2/H^{4/3}$~--- величина гидравлического сопротивления, зависящая от коэффициента трения по Маннингу $n_M$. В качестве базового значения примем $n_M = 0.02$\,с/м$^{1/3}$, характерное для русла Волги в межень \cite{Khrapov-etal:2013,Dyakonova-2014,Pisarev-2012}.
Мы ограничимся рассмотрением 1D-модели при постоянных значениях площади поперечного сечения $A$ и смоченного периметра $P$. Учет зависимостей $A(x)$, $P(x)$ требует обобщенной формы записи для $\lambda$~\cite{Burguete-etal:2006Numerical-boundary-conditions}.

\begin{figure}[tbp]
\begin{center}
\includegraphics[width=350pt,height=165pt]{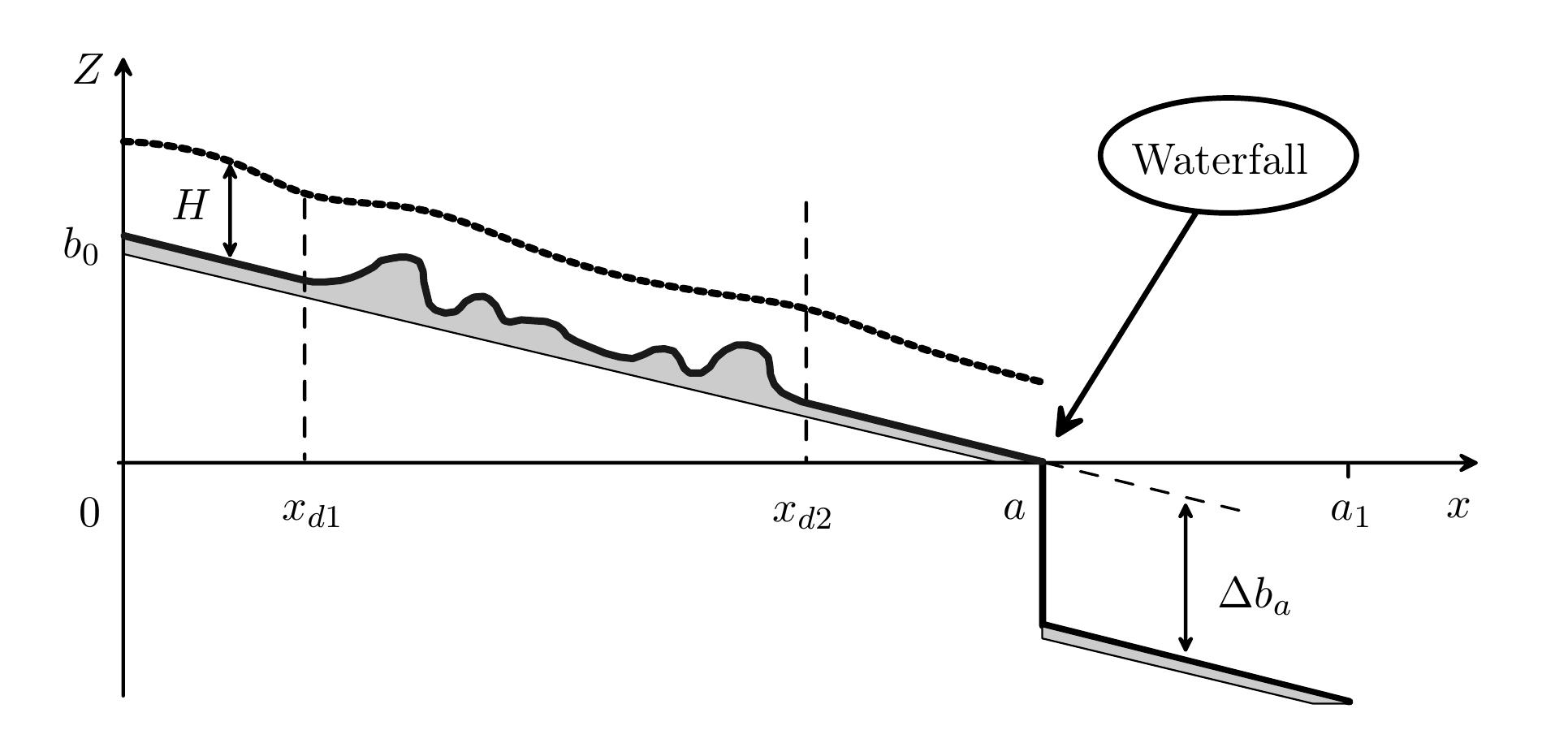}
\end{center}
\caption{Геометрия задачи: поток жидкости стекает по наклонной неоднородной поверхности. В точке $x=a$ дно терпит скачок $\Delta b_a$, создавая условия для водопада}\label{fig-geometry}
\end{figure}

\begin{figure}[tbp]
\begin{center}
\includegraphics[width=250pt,height=251pt]{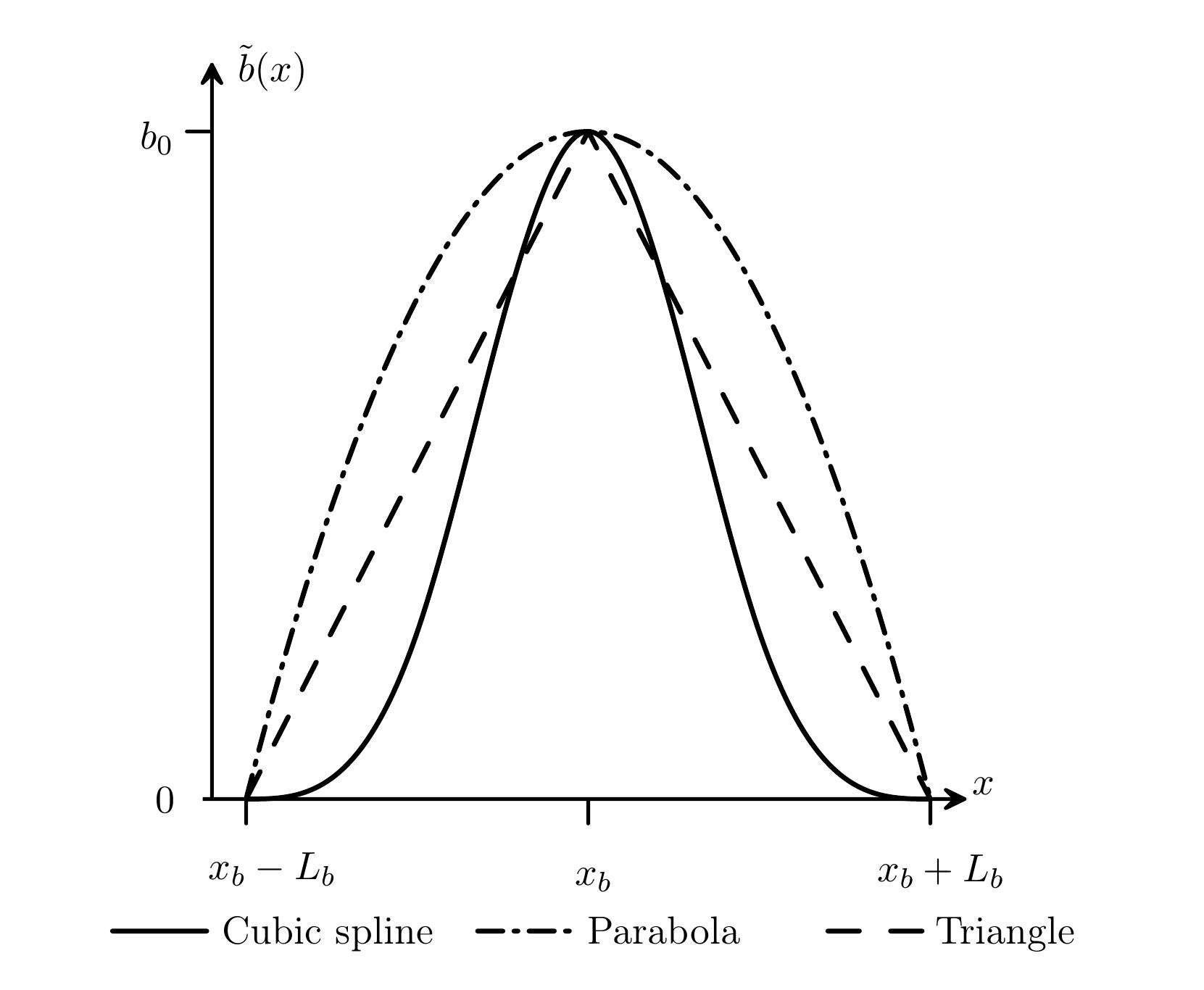}
\end{center}
\caption{Профили локальных симметричных неоднородностей дна}\label{fig-profil-bed}
\end{figure}

На рисунке \ref{fig-geometry} изображена геометрия базовой модели. Функцию рельефа дна представим в виде
\begin{gather}\label{eq-b_all}
b(x)=-\tan(\alpha)\,x \,+ \, \widetilde{b}(x),
\end{gather}
где $\alpha\simeq \tan(\alpha)=(b(0)-b(a))/a$~--- средний угол наклона рельефа дна, приводящий к возникновению руслового течения в заданном направлении. Для гидрологических приложений, связанных с равнинными речными системами и поверхностными стоками, в качестве типичных значений выберем $a=50$\,км и $\tan(\alpha)=0.1$\,м/км.
Вблизи границ в зонах $0 \le x\le x_{d1}$ и $x_{d2} \le x \le a$ считаем дно плоским с углом наклона~$\alpha$, а между точками $x_{d1}$ и $x_{d2}$ разместим различные локальные неоднородности дна $\widetilde{b}(x)$, задавая их в виде
аналитических функций: кубического сплайна
\begin{gather}\label{eq-b_spline}
\widetilde{b}(x) = b_0
\left\{
  \begin{array}{ll}
    1- 6 \,\overline{x}^{\,2} (1 - \overline{x}),  & 0 \le    \overline{x}   \le 0.5; \\
    2 \,(1 - \overline{x})^{3},                    & 0.5 \le      \overline{x}   \le 1; \\
    0,                                             & 1 \le      \overline{x};
  \end{array}
\right.
\end{gather}
параболического и треугольного профилей
\begin{gather}\label{eq-b_p-t}
\widetilde{b}(x) = b_0
\left\{
  \begin{array}{ll}
    1- \overline{x}^{\,k},                & 0 \le    \overline{x}   \le 1; \\
    0,                                    & 1 \le      \overline{x};
  \end{array}
\right.
\end{gather}
где $\overline{x}=|x-x_b|/L_b$, $x_b$~--- положение центра профиля локальной неоднородности дна, $L_b$~--- полуширина профиля, при $k=2$ в (\ref{eq-b_p-t}) имеем параболический профиль, а $k=1$ соответствует треугольному профилю (рис.~\ref{fig-profil-bed}). Набором локальных неоднородностей вида (\ref{eq-b_spline}), (\ref{eq-b_p-t}) с различными значениями параметров $x_b$, $b_0$ и $L_b$ можно моделировать достаточно сложный неоднородный рельеф дна.

%

\section{Численный метод и граничные условия \label{t-p2}}

Для интегрирования уравнений (\ref{eq-H}) и (\ref{eq-q}) воспользуемся лагранжево-эйлеровой численной схемой CSPH-TVD \cite{Khrapov-etal_csph-tvd:2011, Khrapov-etal:2013, Shushkevich-2015}. Данная численная схема является хорошо сбалансированной, консервативной, позволяет сквозным образом рассчитывать динамику воды на неоднородном рельефе дна, содержащем изломы и скачки, а также моделировать гидродинамические течения на реалистичном нерегулярном рельефе при наличии в расчетной области нестационарных границ <<жидкость~--~сухое дно>> и водопадов.

Численное решение будем строить на пространственно-временной сетке $x_{i}=(i-0.5)h$ ($i=\overline{-1,N+2}$) и $t_{n+1}=t_n+\tau_n$ ($n=\overline{0,t_{max}}$), где $h=a/N$~--- пространственный шаг, $N$~--- количество ячеек в расчетной области, {$\tau_n$}~--- временной шаг интегрирования, определяемый из условия устойчивости CSPH-TVD--схемы \cite{Khrapov-etal_csph-tvd:2011, Khrapov-etal:2013}. Величины $H,\,q,\,u$, входящие в уравнения (\ref{eq-H}) и (\ref{eq-q}), при численном интегрировании определяются в узлах пространственно-временной сетки $H_i^n,\,q_i^n\,u_i^n$. Поскольку CSPH-TVD--метод основан на пятиточечной пространственной аппроксимации, то ячейки с индексами $i=-1,0,N+1,N+2$ относятся к <<фиктивным>> ячейкам и для определения значений величин $H,\,q,\,u$ в этих ячейках необходимо задать дополнительные условия на границах расчетной области. Из трех величин $H,\,q,\,u$ только две являются независимыми.

На левой границе $x=0$ наклонной поверхности $b(x)$ будем задавать контролируемый вток воды с расходом $q(x=0,t)=Q(t)$:
\begin{gather}\label{eq-GU-left}
q_{0-j}^n = Q(t_n), \qquad H_{0-j}^n = 2 H_{1-j}^n - H_{2-j}^n, \qquad j=0,1.
\end{gather}
Через правую границу расчетной сетки $x=a$ следует корректно задавать процесс прохождения потока воды, чтобы условия на границе в наименьшей степени влияли на решение в основной части расчетной области. Рассмотрим различные граничные условия и исследуем, как они влияют на характер течения $u(x,t)$, $H(x,t)$.

\begin{flushleft}
{\bf{Свободные граничные условия}}
\end{flushleft}

Значения параметров потока в фиктивных ячейках могут быть определены посредством кусочно-постоянной аппроксимации
\begin{gather}\label{eq-GU-approx0}
    H_{N+1+j}^n = H_{N+j}^n, \quad u_{N+1+j}^n=u_{N+j}^n, \quad j=0,1\
\end{gather}
или кусочно-линейной аппроксимации
\begin{gather}\label{eq-GU-approx1}
    H_{N+1+j}^n = 2H_{N+j}^n-H_{N-1+j}^n, \quad u_{N+1+j}^n=2u_{N+j}^n-u_{N-1+j}^n, \quad j=0,1.
\end{gather}
В соотношениях (\ref{eq-GU-approx0}) и (\ref{eq-GU-approx1}) вместо $H$ или $u$ можно также использовать аппроксимацию величины расхода $q=Hu$.

Граничные условия (\ref{eq-GU-approx0}) и (\ref{eq-GU-approx1}) можно использовать только при сверхкритическом режиме течения с числом Фруда $Fr=u/\sqrt{gH}>1$. При докритическом режиме течения $Fr<1$ данные условия могут оказывать сильное влияние на структуру нестационарного потока из-за интенсивного отражения волн от границы расчетной области, а в случае наличия на границе локальных неоднородностей (отклонения от плоского наклонного профиля) становятся численно неустойчивыми.

Для обеспечения устойчивости граничных условий и уменьшения влияния границ на динамику потока в случае $Fr<1$ применяют инварианты Римана \cite{Yee-etal:1982,Jin-etal:1997,Burguete-etal:2004Implicit-schemes,Burguete-etal:2006Numerical-boundary-conditions},\cite{Cozzolino-2014,Liang-Borthwick-2009}
. При таком подходе одна из величин $H$ или $u$ задается, а другая определяется из условия постоянства инвариантов Римана $u\pm2\sqrt{gH}$ слева и справа от границы.
Пусть, например, значение $H$ в фиктивных ячейках задано, тогда для $u$ имеем
\begin{gather}\label{eq-GU-Rieman-u}
 u_{N+1+j}^n = u_{N+j}^n \pm 2\sqrt{g}\left(\sqrt{H_{N+j}^n}-\sqrt{H_{N+1+j}^n}\right), \qquad j=0,1.
\end{gather}
Аналогично, если задано $u$, то для $H$ выполняется
\begin{gather}\label{eq-GU-Rieman-H}
 H_{N+1+j}^n = \left(\sqrt{H_{N+j}^n} \pm \frac{u_{N+j}^n-u_{N+1+j}^n}{2\sqrt{g}} \right)^2, \qquad j=0,1.
\end{gather}
Знак <<$+$>> в соотношениях (\ref{eq-GU-Rieman-u}) и (\ref{eq-GU-Rieman-H}) соответствует вытекающему потоку, а знак <<$-$>> --- втекающему.

Основная проблема использования граничных условий (\ref{eq-GU-Rieman-u}) и (\ref{eq-GU-Rieman-H}) связана с заданием одной из величин $H$ или $u$ на основе дополнительных соотношений, зависящих от конкретной постановки задачи, структуры течения и рельефа местности. Рассмотрим некоторые варианты определения величины $H$ в фиктивных ячейках для условия (\ref{eq-GU-Rieman-u}):
\begin{gather}\label{eq-GU-Rieman-u-H}
  H_{N+1+j}^n = H_{N+j}^n + b_{N+j} - b_{N+1+j} - \tan(\alpha_\eta) h\,, \qquad j=0,1\,,
\end{gather}
где $\alpha_\eta$~--- свободный параметр, характеризующий угол наклона уровня воды $\eta$ на правой границе расчетной области.
Значение $\alpha_\eta$ зависит от профиля рельефа дна в окрестности границы и выбирается таким образом, чтобы обеспечить устойчивость граничных условий и минимизировать влияние границы на численное решение. Так, например, для плоского дна $b_{N+j} - b_{N+1+j}=\tan(\alpha) h$ (см.~рис.~\ref{fig-geometry}) можно принять $\alpha_\eta=\alpha$. В этом случае условие (\ref{eq-GU-Rieman-H}) будет соответствовать условию для $H$ в соотношении (\ref{eq-GU-approx0}). При наличии на границе локальных неоднородностей дна $\widetilde{b}(x)$ для руслового течения необходимо выбирать $\alpha_\eta\geq\alpha$.

 Другим подходом, способным обеспечить устойчивость граничных условий (\ref{eq-GU-Rieman-u})--(\ref{eq-GU-Rieman-u-H}),  является модификация рельефа дна в окрестности границы
\begin{gather}\label{eq-GU-Rieman-u-b}
  b_{N+1+j} = b_{N+j} - \tan(\widetilde{\alpha}) h, \qquad j=0,1,
\end{gather}
где $\widetilde{\alpha}$~--- свободный параметр, характеризующий угол наклона дна на правой границе расчетной области. Такой подход требует сильных изменений рельефа для двумерных течений вблизи границ.

Граничные условия (\ref{eq-GU-approx1}) будем называть условиями типа I, а с применением инвариантов Римана --- условиями типа II.

\begin{flushleft}
{\bf{Граничные условия типа <<поглощение>>}}
\end{flushleft}

Наиболее простым граничным условием, обеспечивающим устойчивый расчет и исключающим проникновение жидкости в расчетную область  через правую границу, является условие типа <<поглощение>>:
\begin{gather}\label{eq-GU-absorption}
    H_{N+j}^n = 0, \qquad u_{N+j}^n=0, \qquad j=1,2.
\end{gather}
 Граничные условия типа <<поглощение>> (\ref{eq-GU-absorption}) соответствуют наличию водопада в точке $x=a$ (см. рис.\,\ref{fig-geometry}). Если выполняется условие $H_{N+j}^n<\Delta b_a$, то течение в области $x>a$ не оказывает влияния на решение до водопада.

Ниже рассмотрим вопрос о применимости и влиянии на динамику течения граничных условий типа <<поглощение>> (тип III). В предельном случае $\alpha_\eta\rightarrow\pi/2$ получаем переход к условию (\ref{eq-GU-absorption}) с учетом $H\geq0$.

\section{Влияние граничных условий на стационарный поток}\label{GU-stationary-flow}

\begin{flushleft}
{\bf{Точное решение}}
\end{flushleft}

При стационарном течении $\partial H /\partial t\equiv\partial q /\partial t\equiv 0$ уравнения (\ref{eq-H}) и (\ref{eq-q}) принимают вид
\begin{gather}\label{eq-equlib-1}
\frac{dq}{dx}=0\, \quad \text{или} \quad q=\textrm{const},
\end{gather}
\begin{gather}\label{eq-equlib-2}
{d\over dx}\left(\frac{u^2}{2}+g\eta\right) = -  \frac{\lambda}{2}u |u|.
\end{gather}
Заметим, что величина ${u^2}/{2}+g\eta$ в (\ref{eq-equlib-2}) является постоянной интеграла Бернулли и под действием сил придонного трения изменяется вдоль линий тока из-за диссипации энергии потока.

С учетом (\ref{eq-b_all}), (\ref{eq-equlib-1}) и выражения для коэффициента гидравлического сопротивления $\lambda$ уравнение (\ref{eq-equlib-2}) может быть представлено в виде
\begin{gather}\label{eq-equlib-dH}
{d H\over dx} = \frac{\displaystyle{-\frac{d\widetilde{b}}{dx}} + \tan(\alpha) \left[1 - (H_*/H)^{10/3}\right]}{1-Fr^2},
\end{gather}
где $\displaystyle H_* = \left(\frac{q^2 n_M^2}{\tan(\alpha)}\right)^{3/10}$, $Fr$~--- число Фруда ($Fr=u/\sqrt{gH}$). При стационарном течении на плоском наклонном дне
с $\alpha=\textrm{const}$ и $n_M=\textrm{const}$ решением уравнения (\ref{eq-equlib-dH}) является $H(x)=H_*=\textrm{const}$. И в этом случае для числа Фруда имеем $\displaystyle Fr_* = \frac{q}{\sqrt{gH_*^3}}=\frac{q^{1/10}\tan^{9/20}(\alpha)}{g^{1/2}n_M^{9/10}}$.

Подчеркнем, что уравнение (\ref{eq-equlib-dH}) можно использовать для построения стационарных решений только для гладких функций $\widetilde{b}$ и в отсутствии критических точек $Fr=1$.

Ниже ограничимся рассмотрением решений (\ref{eq-equlib-dH}) $Fr>1$, либо $Fr<1$.

Решение уравнения (\ref{eq-equlib-dH}) для стационарного потока будем называть точными,
а соответствующее решение системы уравнений (\ref{eq-H}), (\ref{eq-q}) на основе CSPH-TVD метода для установившегося течения (в пределе $t\rightarrow\infty$) --- численным.
 Для получения установившегося течения при численном моделировании достаточно задать на левой границе постоянный приток жидкости
$Q = \textrm{const}$ в (\ref{eq-GU-left}).
В качестве базовых параметров в численной модели примем $Q = 20$ м$^2$/с, $N=8000$.
Начальные условия для системы уравнений (\ref{eq-H}), (\ref{eq-q}) будем задавать в виде пространственно-однородных распределений $H(x,t=0)=H_*$, $q(x,t=0)=Q$.
 Течения будем называть установившимися (стационарными) на временах $t\geq t_s$ при выполнении $\max_i|H_i^{n+1}/H_i^n\,-\,1|<10^{-10}$, что  достигается в численной модели при $t_s\gtrsim100$~ч для выбранных значений $Q$ и $a$.

\begin{figure}[tbp]
\begin{center}
\includegraphics[width=250pt,height=261pt]{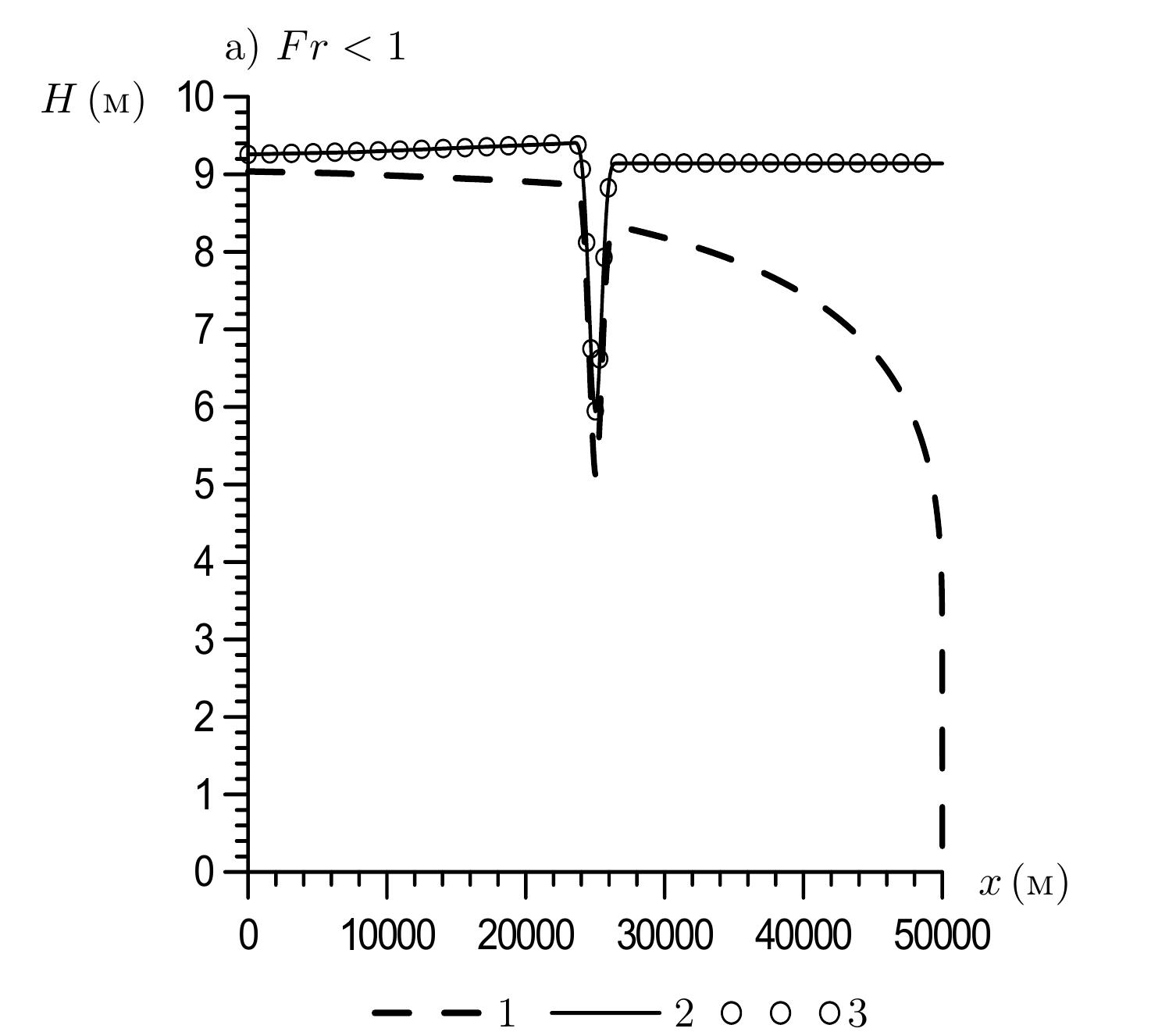}
\includegraphics[width=250pt,height=258pt]{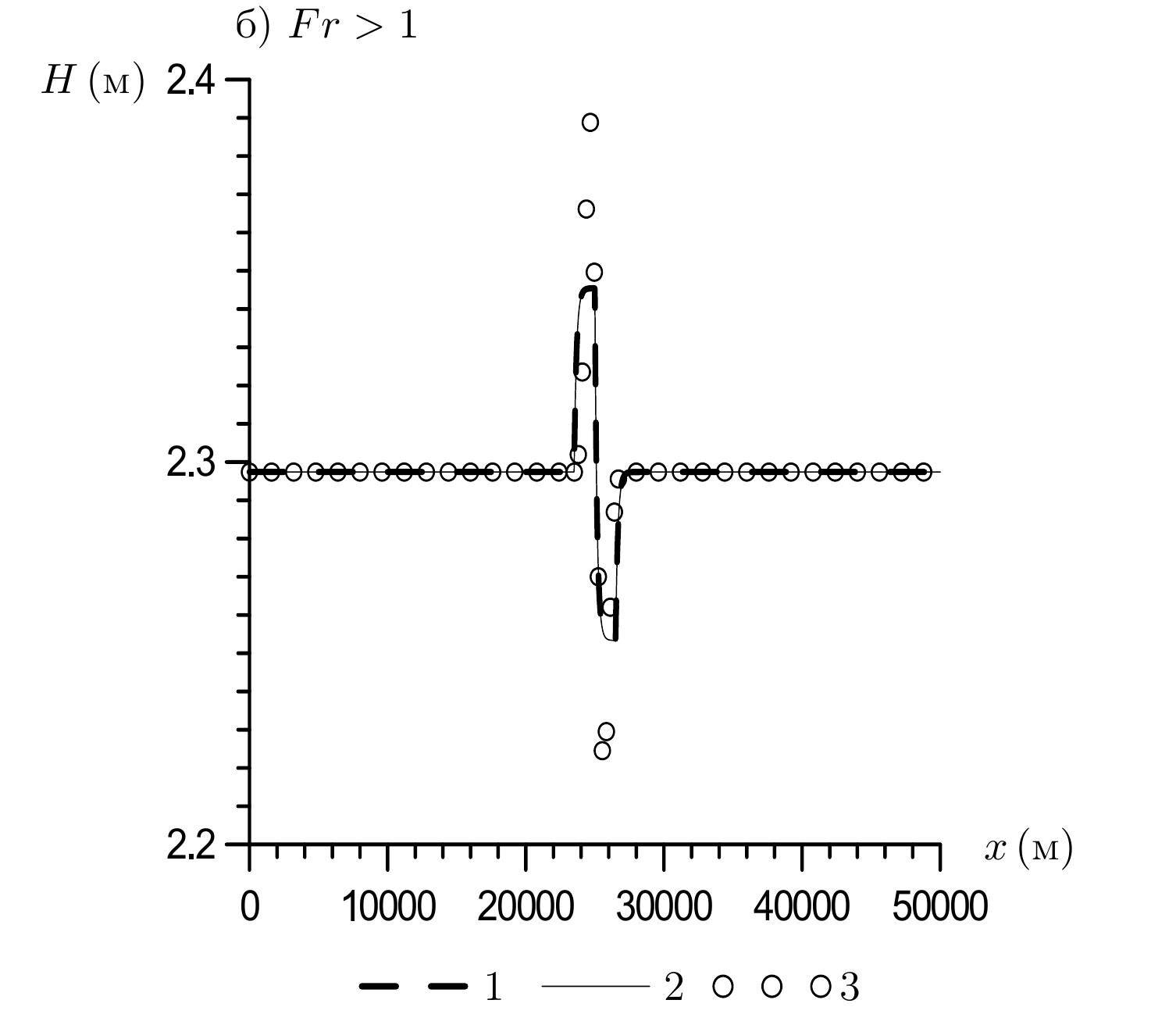}
\end{center}
\caption{Распределения глубины $H(x)$ для трех моделей в случае локальной неоднородности дна (см. рис.~\ref{fig-profil-bed}) с параметрами $x_b=2500$ м, $L_b=1500$ м, $b_0=7$ м для кубического сплайна (\ref{eq-b_spline}): линия 1~--- граничное условие типа III; линия 2~--- типа I; линия 3~--- точное решение (\ref{eq-equlib-dH}). а~--- случай докритического течения, б~--- случай сверхкритического течения.}\label{fig-exact-numerical}
\end{figure}

На рисунке \ref{fig-exact-numerical} изображены точные и численные решения для стационарного потока воды на неоднородном дне при различных режимах течения. Относительное отклонение численного решения от точного на интервале $(0;2000)$ в случае докритического течения (см.~\ref{fig-exact-numerical}\,\textit{а}) не превышает $0.1\%$ для граничных условий типа I, а для граничных условий типа III погрешность существенно возрастает до $10\%\div30\%$. В случае сверхкритического течения (см.~\ref{fig-exact-numerical}\,\textit{б}) относительная погрешность в рассматриваемой области не превышает $0.01\%$ для обоих типов граничных условий.
С увеличением $N$ погрешность уменьшается пропорционально $1/N^2$, что свидетельствует о квадратичной сходимости численного метода \cite{Khrapov-etal_csph-tvd:2011}.
 Отметим, что численные решения для установившихся течений на плоском дне со свободными граничными условиями (\ref{eq-GU-approx0}), (\ref{eq-GU-approx1}), (\ref{eq-GU-Rieman-u}) и (\ref{eq-GU-Rieman-u-H}) тождественно совпадают с точными в пределах точности моделирования $10^{-15}$ (для 8 байтовых чисел).

\section{Плоское дно}\label{bottom}

\begin{figure}[tbp]
\begin{center}
\includegraphics[width=225pt,height=166pt]{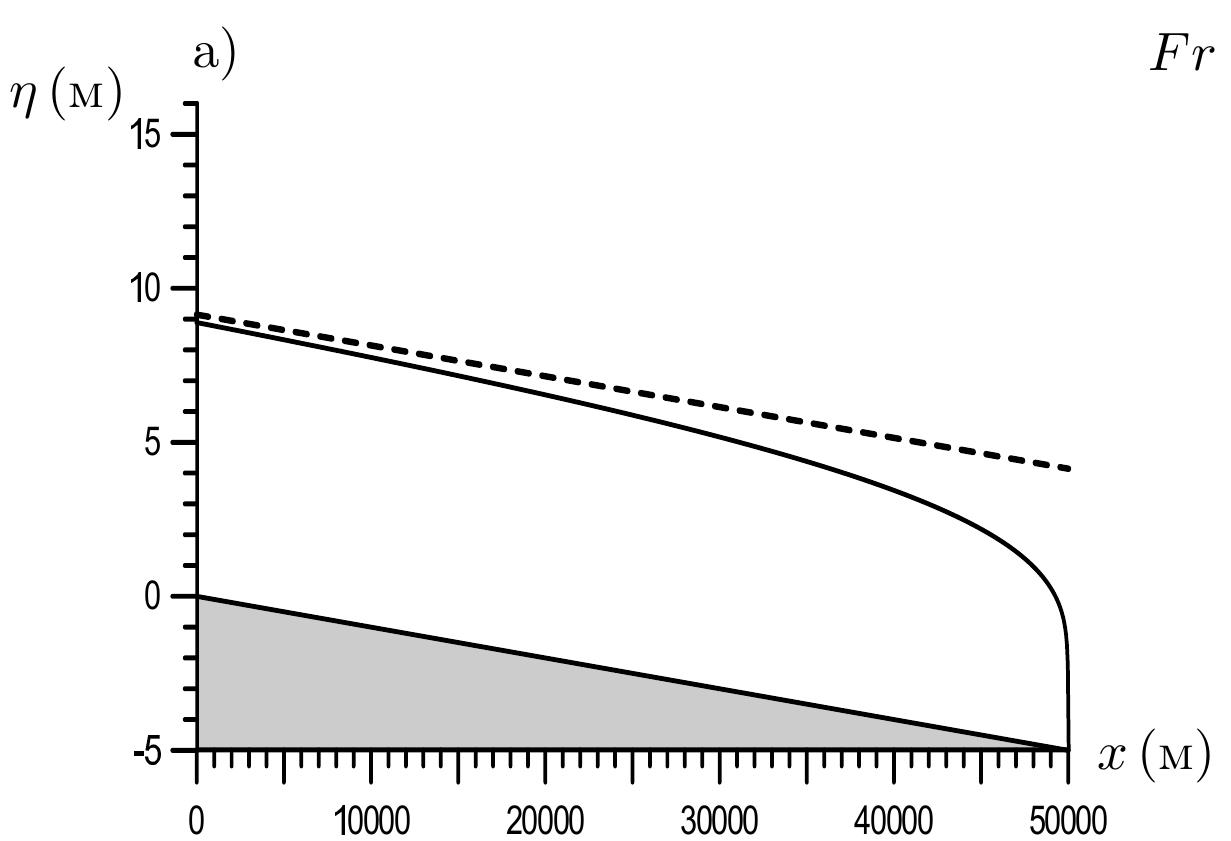}
\includegraphics[width=225pt,height=166pt]{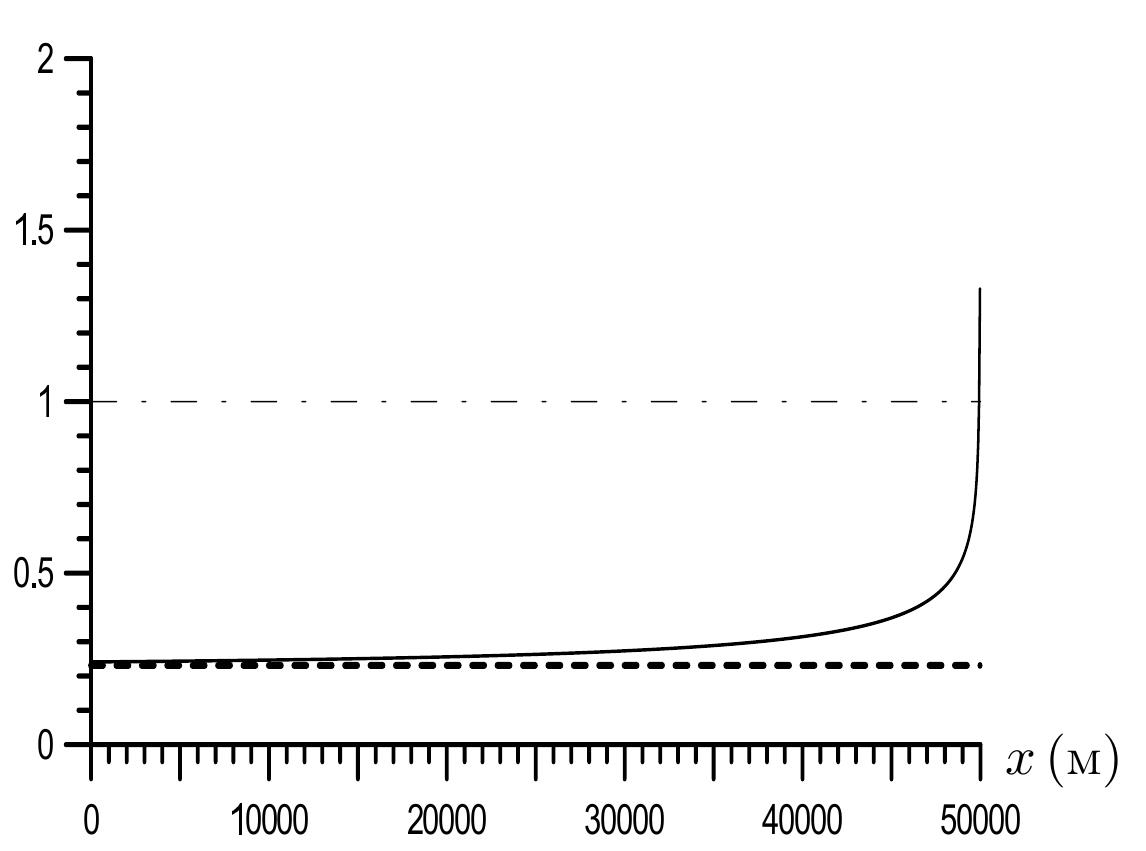}
\includegraphics[width=225pt,height=165pt]{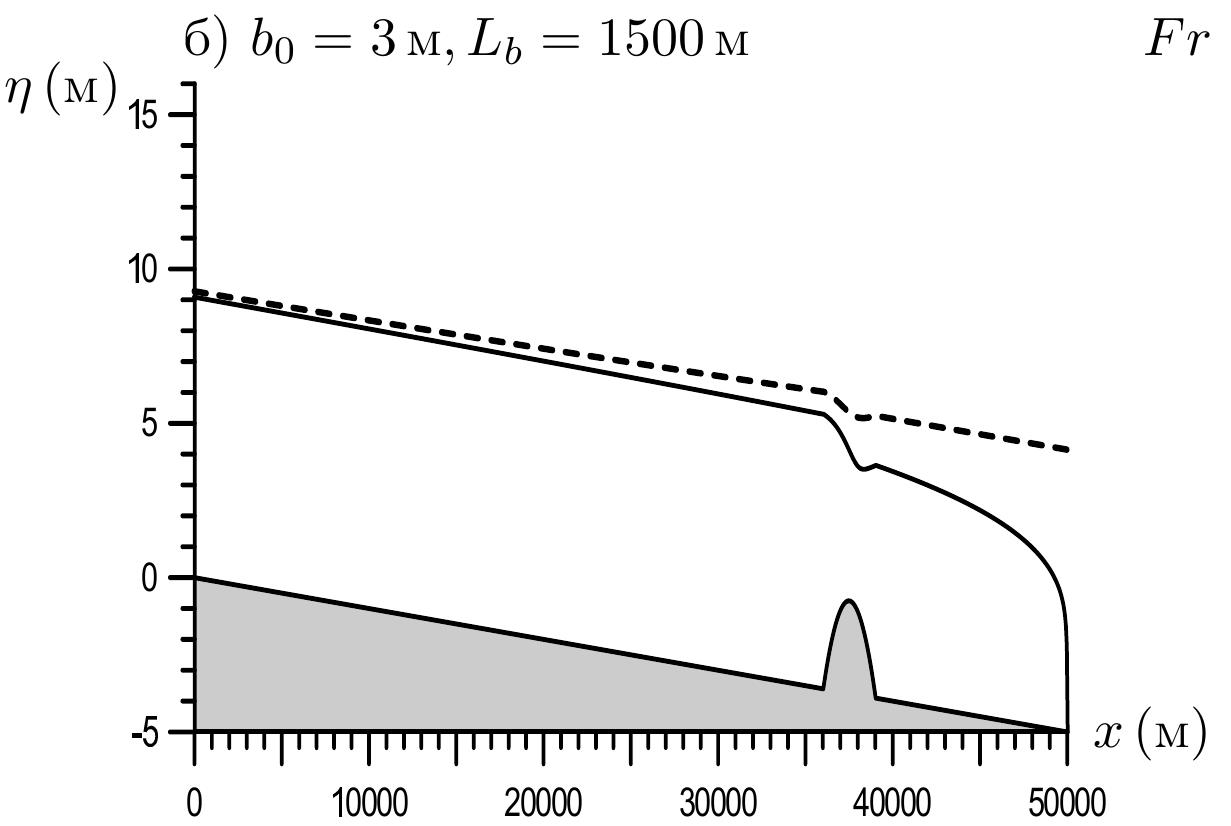}
\includegraphics[width=225pt,height=165pt]{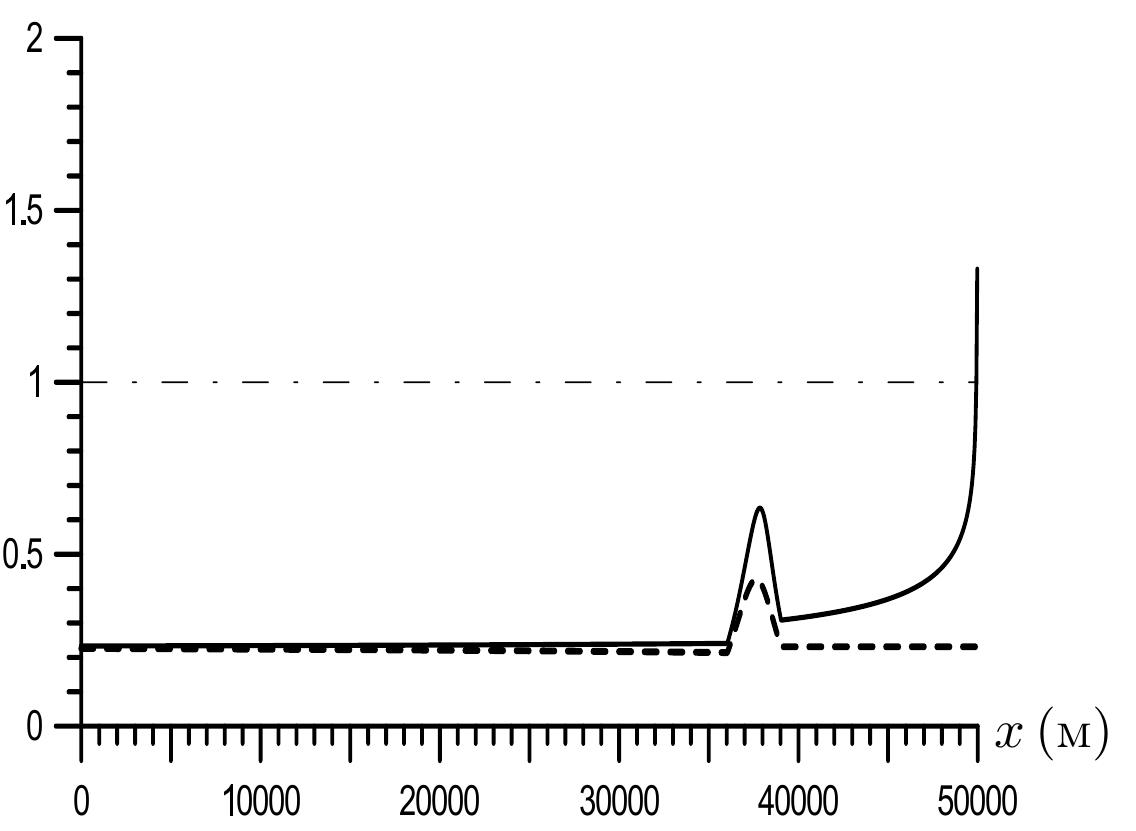}
\includegraphics[width=225pt,height=166pt]{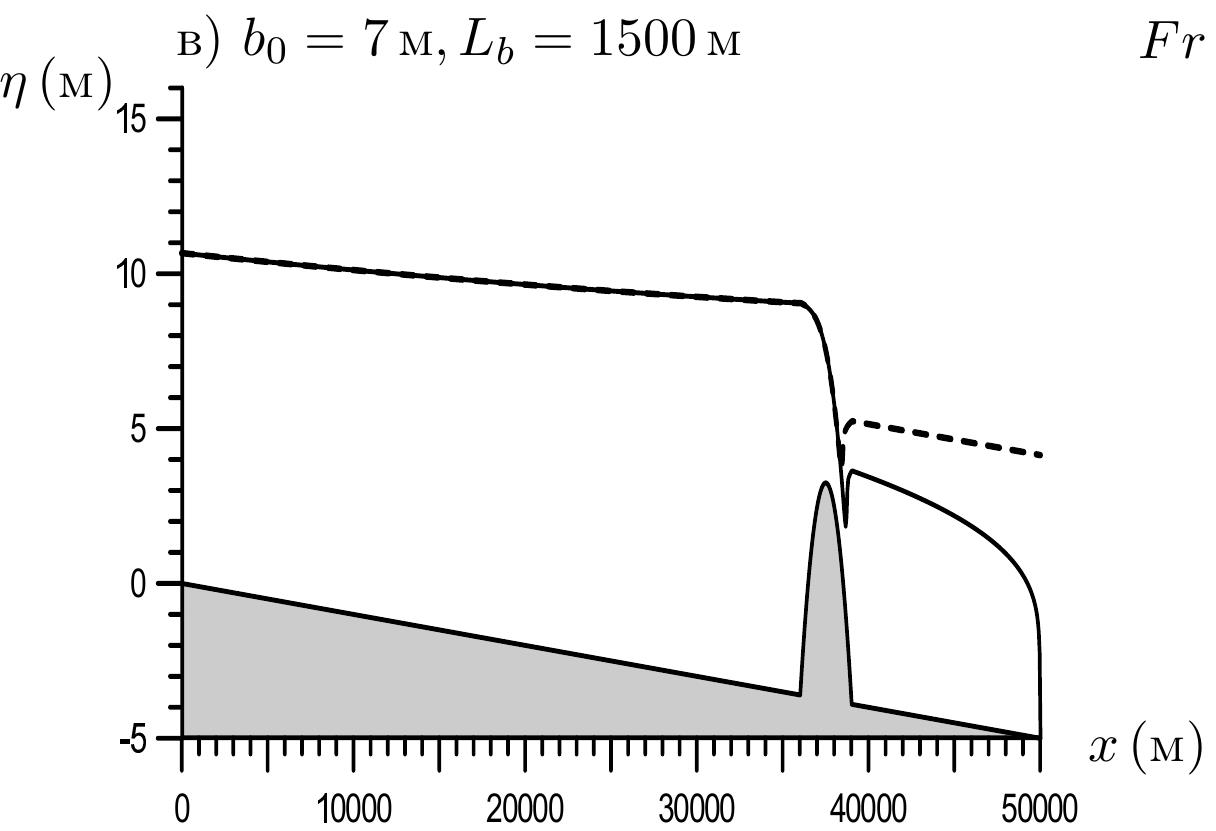}
\includegraphics[width=225pt,height=166pt]{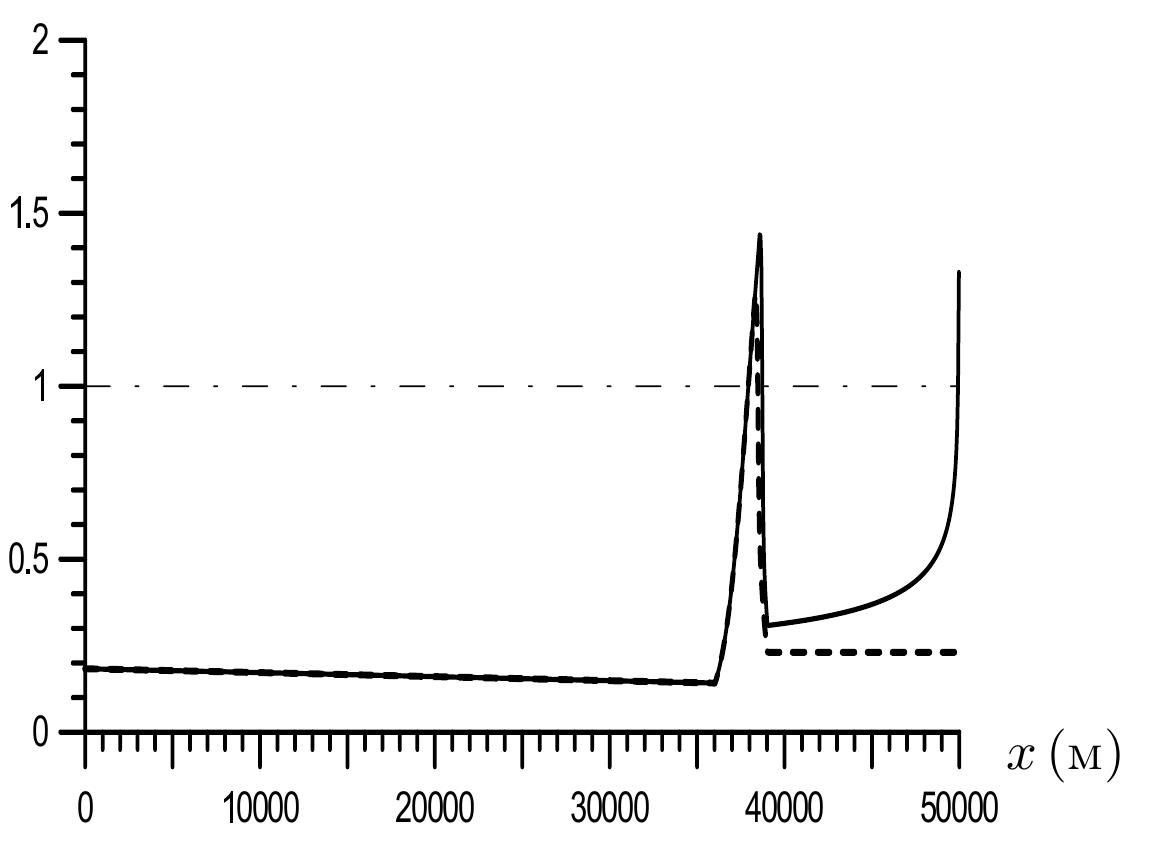}
\includegraphics[width=225pt,height=167pt]{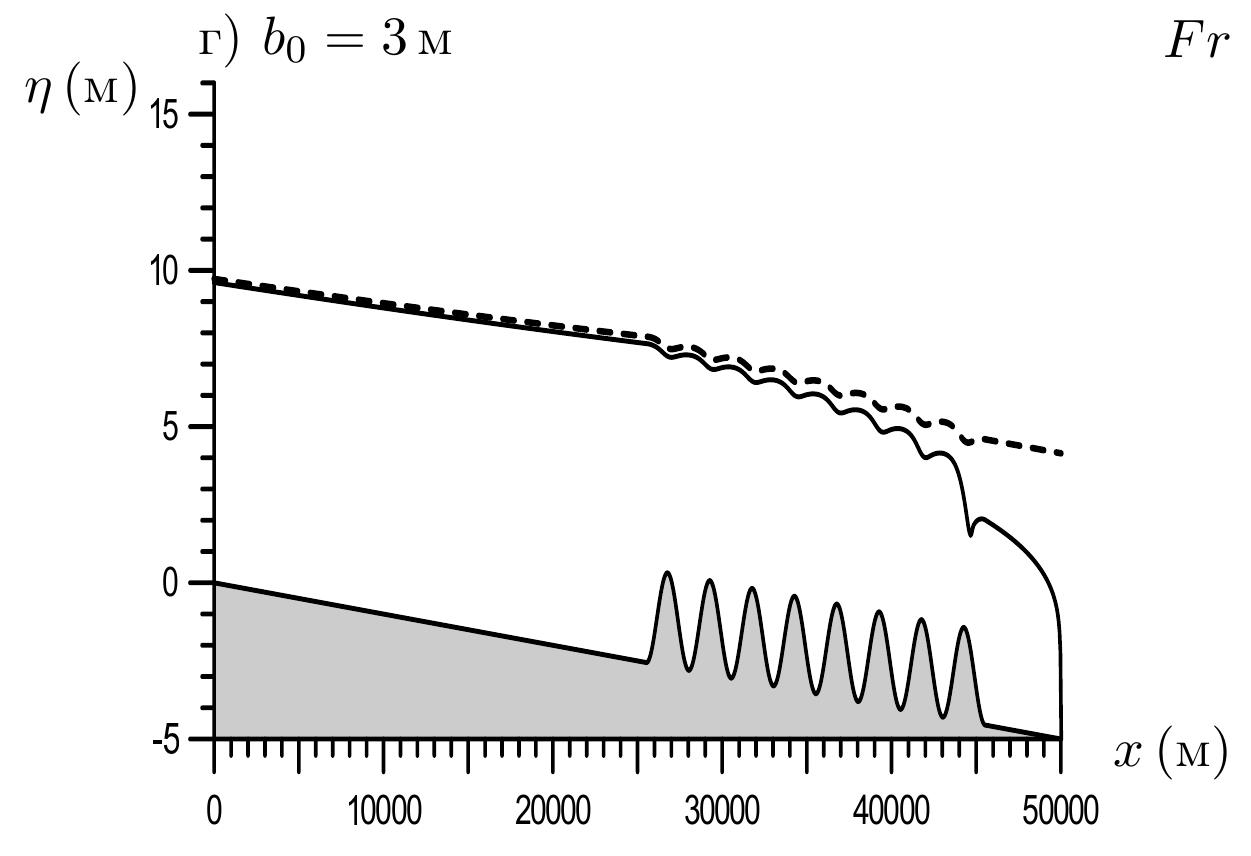}
\includegraphics[width=225pt,height=167pt]{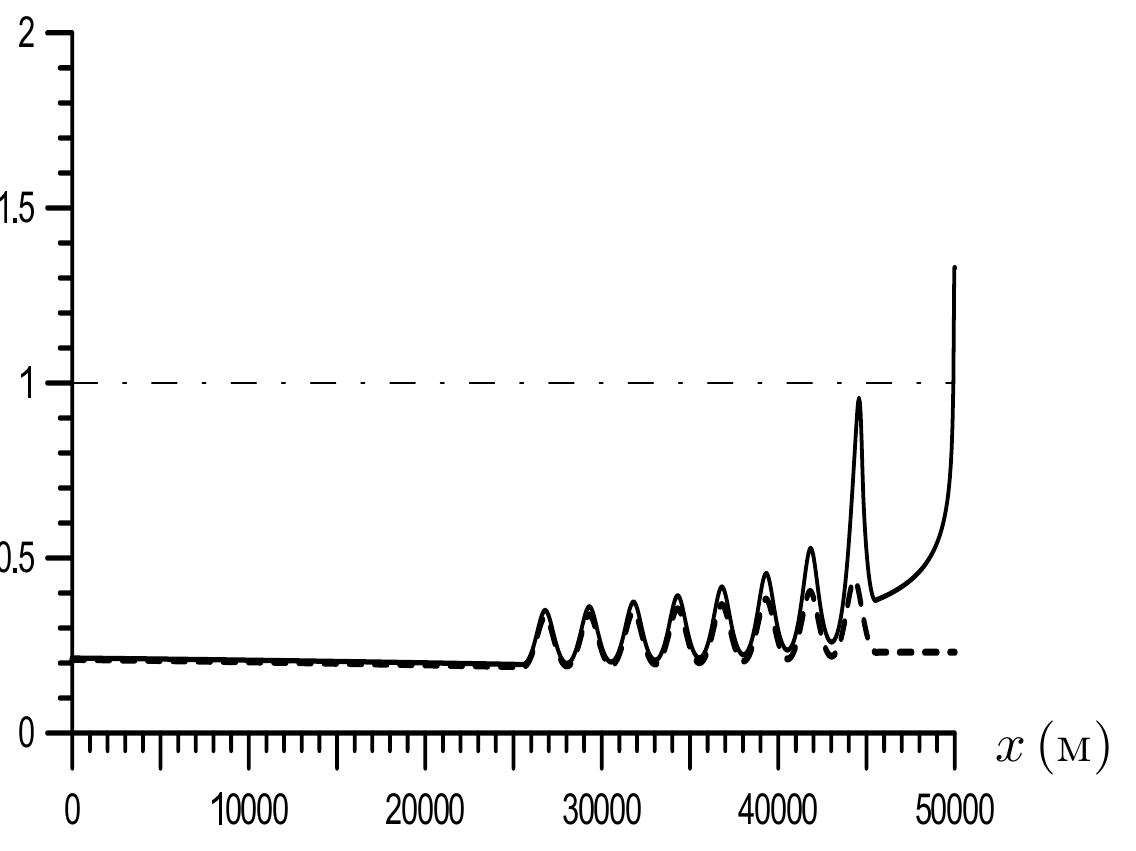}
\end{center}
\caption{Распределения уровней воды $\eta(x)$ и числа Фруда $Fr(x)$ (сплошная линия --- граничное условие типа III, штриховая линия --- граничное условие типа I): \textit{а}) гладкое дно; \textit{б}) возникновение области со сверхкритическим потоком на неоднородности; \textit{в}) докритический поток при наличии одной неоднородности; \textit{г}) докритический поток при наличии системы неоднородностей дна}\label{fig-fig02}
\end{figure}

 Рассмотрим течение с постоянным уклоном дна во всей расчетной области ${\alpha=-db}/{dx}=\textrm{const}>0$.
 Сравним решения при использовании различных граничных условий.
Граничные условия при использовании инвариантов Римана (\ref{eq-GU-Rieman-u}) и линейной аппроксимации (\ref{eq-GU-approx1}) совпадают для всей расчетной области (рис.~\ref{fig-fig02}\,\textit{а}).

  Для гладкого дна влияние граничного условия типа поглощения (\ref{eq-GU-absorption}) простирается на десятки километров выше по течению (см. рис.~\ref{fig-fig02}\,\textit{а}). При удалении от правой границы разность $\eta^{(I)} -\eta^{(III)}$ уменьшается достаточно медленно, и даже на расстоянии 20 км различие составляет около 1 метра ($\simeq 10$\,\%).

 Варьирование угла уклона $\alpha$ в пределах от $0.04$\,м/км (случай реки Обь) до $4$\,м/км (случай реки Терека) и коэффициента шероховатости в пределах $n_M=0 \div 0.1$ сохраняет данный результат.

\section{Дно с локальными неоднородностями}

Неоднородности дна будем располагать в области $1/2< x/a \le 1$, а на однородном дне $0\le x/a < 1/2$ будем рассчитывать относительную интегральную разницу между решениями
\begin{gather}\label{eq-epsilon}
    \varepsilon = \frac{2}{a} \int\limits_{0}^{a/2} \frac{\left| \eta^{(I)} -\eta^{(III)} \right|}{\eta^{(I)}} \,dx \,.
\end{gather}

Рассмотрим влияние одиночных препятствий (рис.\,\ref{fig-fig02}\,\textit{б,в}) и сложных неоднородностей дна (рис.\,\ref{fig-fig02}\,\textit{г}) на стационарные решения с использованием различных граничных условий, ограничившись достаточно крупномасштабными неоднородностями с относительной высотой $b_0/H\simeq 0.1\div 0.7$ и горизонтальной шкалой $L_b\simeq (10\div 400)H\gg H$. Неоднородность дна может существенно изменять структуру потока выше по течению. Появление зоны сверхкритического течения $Fr>1$ и гидравлического скачка качественно изменяют ситуацию выше по течению (рис.~\ref{fig-fig02}\,\textit{в}), где решения $\eta^{(III)}(x)$ и $\eta^{(I)}(x)$ практически совпадают.
Существует набор параметров, характеризующих локальные неоднородности дна, обеспечивающие возникновение сверхкритической зоны ($Fr>1$).
Во всех случаях ниже по течению от зоны неоднородности дна разность $\left|\eta^{(I)} -\eta^{(III)}\right|$ велика (см. область $x=40-50$\,км на рис.~\ref{fig-fig02}\,б,\,{в}).
А выше по течению от этих неоднородностей величина $\left|\eta^{(I)} -\eta^{(III)}\right|$ уменьшается и в случае докритического течения (рис.~\ref{fig-fig02}\,\textit{б, г}).
 Причем более сложный характер неоднородного рельефа существенно ослабляет влияние граничных условий выше по течению (рис.~\ref{fig-fig02}\,\textit{г}). Рис.~\ref{fig-fig02}\,\textit{г} наглядно демонстрирует уменьшение влияния поглощающих граничных условий при наличии нескольких неоднородностей дна даже без формирования сверхкритического режима.

\begin{figure}[tbp]
\begin{center}
\includegraphics[width=250pt,height=259pt]{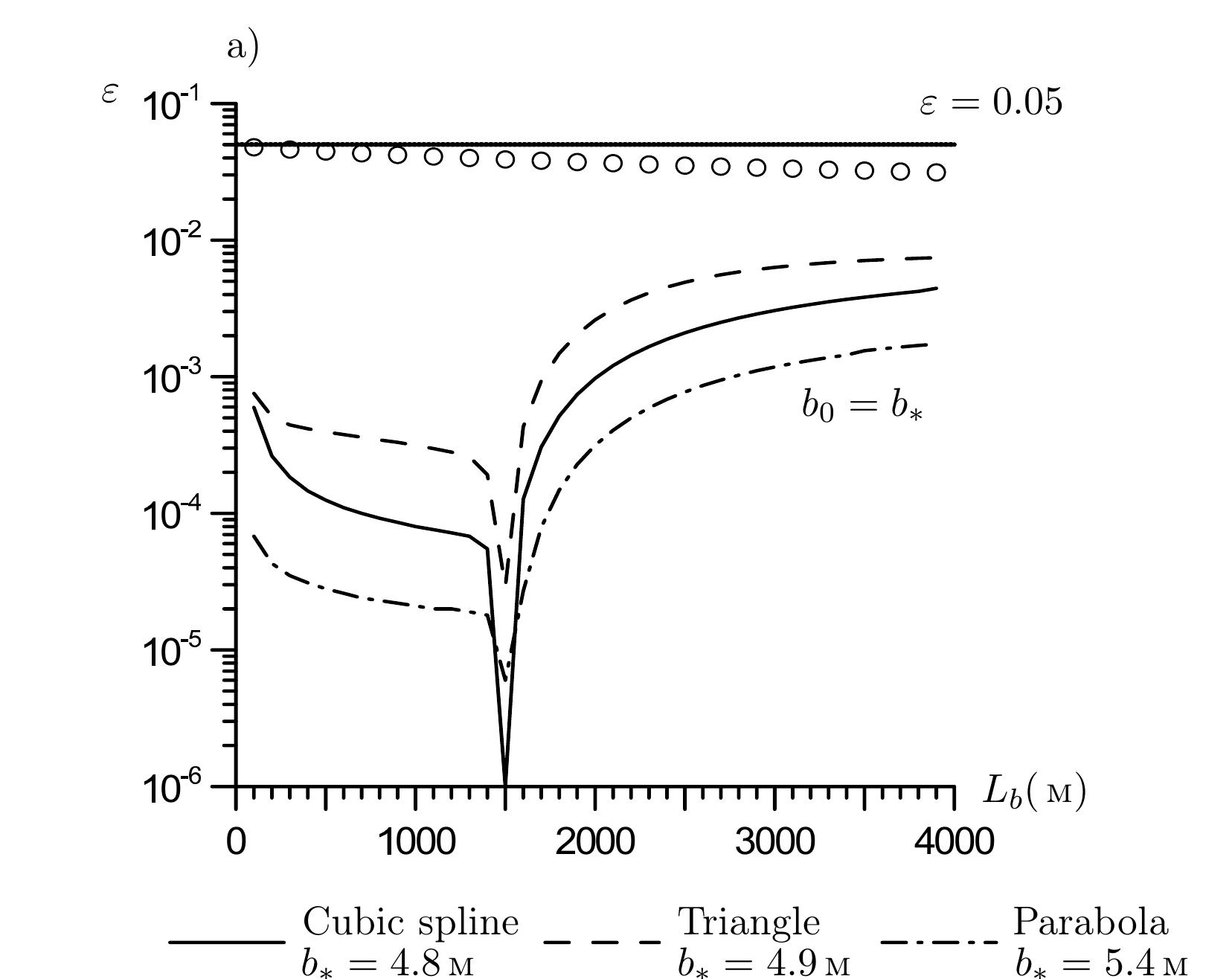}
\includegraphics[width=250pt,height=262pt]{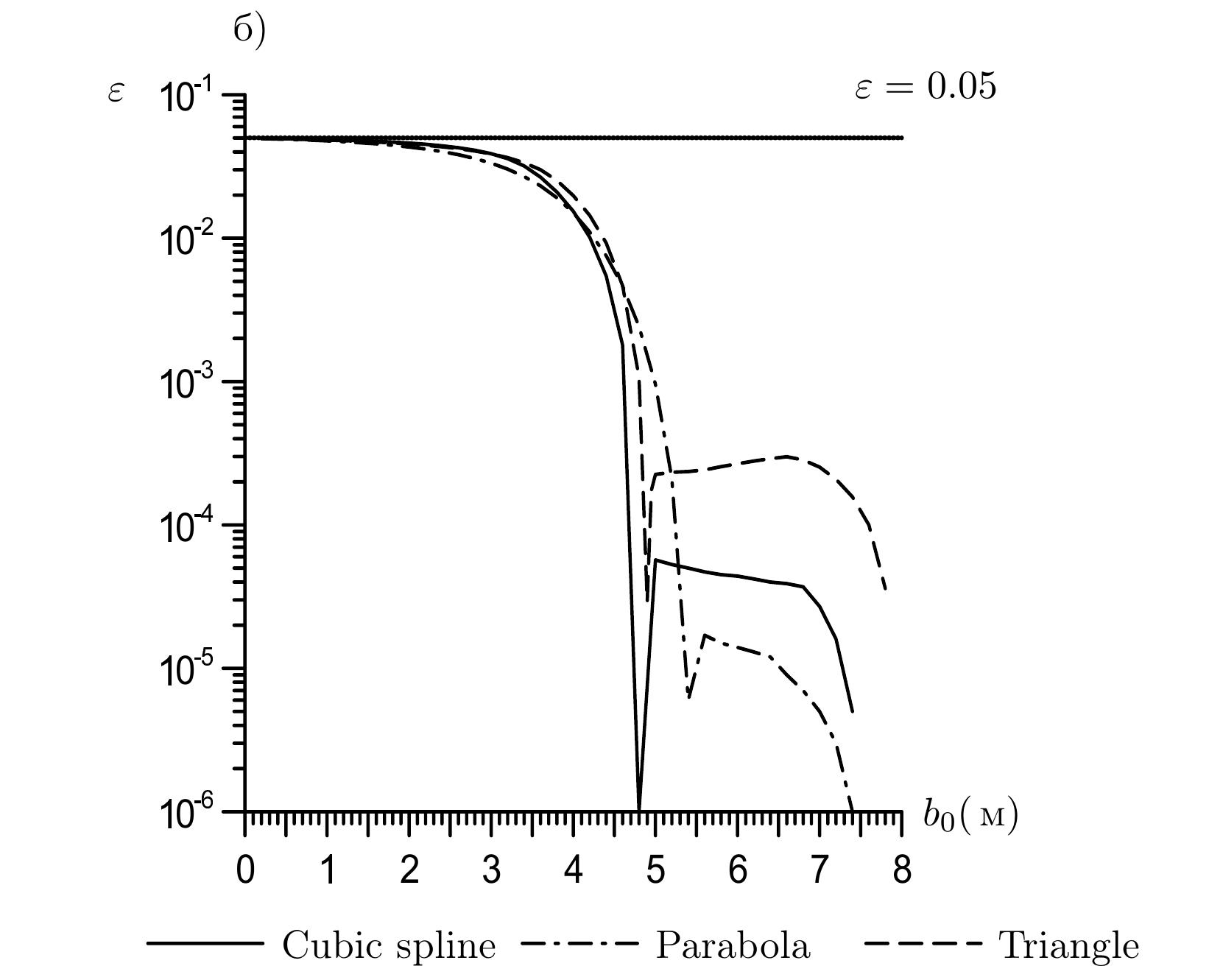}
\end{center}
\caption{Влияние параметров неоднородности дна на общую структуру течения. Точечная линия соответствует течению на плоском дне, а кружками обозначено докритическое течение для треугольного профиля с $b_0=3$. {а}) зависимость $\varepsilon(L_b)$;
{б}) $\varepsilon(b_0)$ для фиксированного $L_b=1500$\,м.
}\label{fig-fig03x}
\end{figure}

 Рассмотрим влияние параметров $b_0$ и $L_b$, характеризующих неоднородность различного вида, на величину $\varepsilon$ (\ref{eq-epsilon}).
На рисунке \ref{fig-fig03x} показаны зависимости $\varepsilon(L_b)$, $\varepsilon(b_0)$ для трех характерных форм возмущений дна в виде
(\ref{eq-b_spline}) и (\ref{eq-b_p-t}) (см. рис.~\ref{fig-profil-bed}). Для выбранных значений расчетной области $a=50$\,км и величины стока $Q=20$\,м$^2/$с ($H\simeq 10$\,м) на гладком дне имеем $\varepsilon=0.05$ (см. \ref{fig-exact-numerical}\,\textit{а} и предел $b_0\rightarrow 0$ на рис.~\ref{fig-fig03x}). При увеличении $b_0$ имеем уменьшение расхождения между решениями. Причем при определенных $b_0$ формируются критические течение, сопровождающее резким уменьшением $\varepsilon$ (см. область $b_0\simeq 5$\,м в случае $L_b=1500$\,м). Этот результат слабо зависит от формы неоднородности. При больших амплитудах $b_0$ расхождение между решениями находится в пределах $\lesssim 10^{-5}\div 10^{-4}$.

\begin{figure}[tbp]
\begin{center}
\includegraphics[width=450pt,height=166pt]{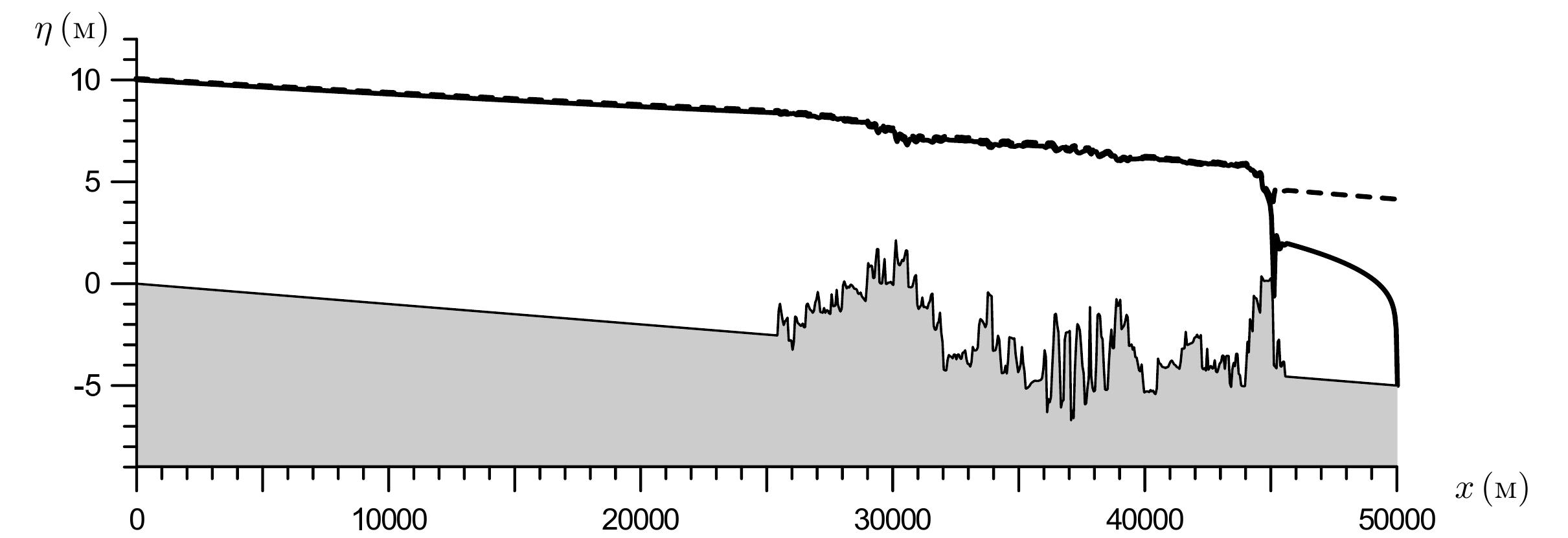}
\includegraphics[width=450pt,height=166pt]{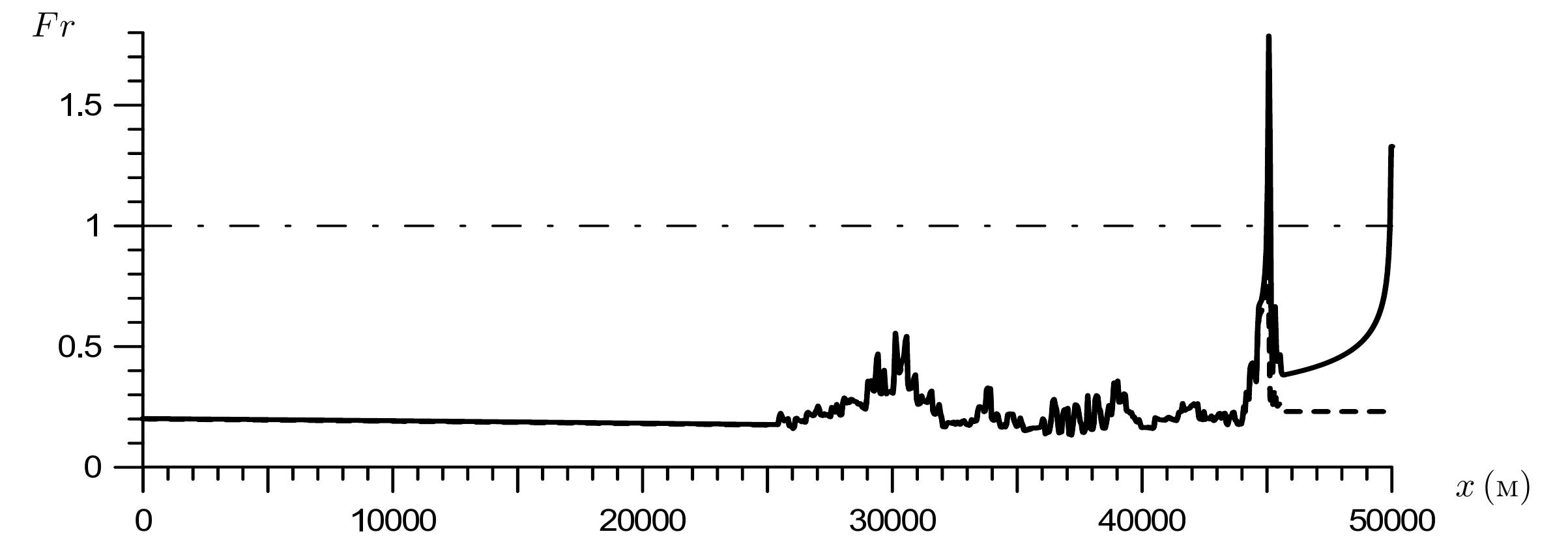}
\end{center}
\caption{Распределения уровней воды $\eta(x)$ и числа Фруда $Fr(x)$ (сплошная линия --- граничное условие типа III, штриховая линия --- граничное условие типа I) для модели с реальным профилем дна для участка р. Волга на интервале $x=(25-46)$\,км.}\label{fig-fig02x}
\end{figure}

Реальный рельеф дна существенно отличается от модельного на рис. \ref{fig-fig02}\,\textit{г}.
В качестве реального рельефа дна выберем 20-ти километровый участок русла Волги вблизи Светлого Яра (примерно 50 км ниже плотины Волжской ГЭС), который расположим внутри расчетной области. На рис.\,\ref{fig-fig02x} приведены результаты расчетов. Мы видим очень хорошее совпадение решений при использовании линейной аппроксимации на границе и условий в виде водопада уже при $x<45$\,км, причем граничные условия III дают зону критического течения вблизи $x\simeq 45$\,км, а граничные условия I сохраняют докритический режим на всей расчетной области.

\begin{flushleft}
{\bf{Заключение}}
\end{flushleft}

При численном моделировании поверхностного потока мелкой воды в случае сложного неоднородного рельефа дна использование граничных условий типа поглощения (водопад) дает вполне адекватные результаты для широкого круга задач.
Наличие неоднородностей дна (отмели, косы, ямы) качественно меняет характер течения по сравнению с плоским дном. Размеры области, на которую оказывают влияние граничные условия, существенно уменьшаются  выше по течению. Выбор того или иного граничного условия влияет на структуру течения только в непосредственной близости от границы и разница между решениями становится исчезающе малой выше по течению.
По-видимому, причина связана с эффективным отражением волны разрежения из-за частичных отражений от любых неоднородностей дна. На это указывает то, что влияние неоднородности рельефа $b(x)$ на структуру течения в виде ямы оказывается меньше, чем в форме поднятия дна русла.
Если на неоднородности возникает сверхкритический поток с гидравлическим скачком, то это в наибольшей степени ослабляет влияние самой границы на структуру потока выше по течению.

Реальные русла характеризуются большим количеством неоднородностей рельефа дна на самых различных масштабах, что позволяет использовать более простые граничные условия в виде поглощения (типа водопад), обеспечивая погрешность лучше 1\,\% уже на расстоянии несколько километров от границы.
В случае двумерных течений дополнительными факторами неоднородности являются изменения ширины и формы поперечного профиля русла, изгибы и повороты, вплоть до меандрирования русла. Эти особенности вносят дополнительный вклад в ослабление влияния границы.
\vspace{3ex}

\small

\begin{center}
REFERENCES
\end{center}
\setlength{\leftmargini}{1.8em}
\begin{enumerate}
\setlength{\itemsep}{0em}
\parskip=0pt

\item
 Shokin Yu.I., Fedotova Z.I., Khakimzyanov G.S {\it Ierarkhiya nelineinykh modelei gidrodinamiki dlinnykh poverkhnostnykh voln}, {\it Doklady Akademii Nauk}, 2015, vol. 462, no. 2, pp. 168--172.

\item
 Fedotova Z.I., Khakimzyanov G.S Nonlinear dispersive shallow water equationsfor a non-stationary bottom, {\it Vychisl. Tekhn.}, 2008, vol. 13, no. 4, pp. 114--126 (in Russian).

\item
 Green A.E., Naghdi P.M. A derivation of equations for wave propagation in water of variable depth, {\it Journal of Fluid Mechanics}, 1976, vol. 78, no. 02, pp. 237--246.

\item
 Bautin S.P., Deryabin S.L. {\it Issledovanie nachal'no-kraevoi zadachi dlya sistemy uravnenii Grina-Nagdi}, {\it Vestn. Ural. Gos. Univ. Put. Soob.}, 2012, no. 1 (13), pp. 4--13.

\item
 Pelinovskii E.N. {\it Gidrodinamika voln tsunami}, Nizhnii Novgorod:  Institut Prikladnoi Fiziki Ross. Akad. Nauk, 1996, 276 p.

\item
 Bautin S.P., Deryabin S.L., Sommer A.F., Khakimzyanov G.S. Investigation of solutions of the shallow water equations in the vicinity of the mobile run-up line, {\it Vychisl. Tekhnol.}, 2010, vol. 15, no. 6, pp. 19--41 (in Russian).

\item
 Prokof'ev V.A. {\it Utochnenie modeli melkoi vody na osnove spektral'nogog predstavleniya profilya skorosti po glubine}, {\it Izvestiya Vserossiiskogo Nachno-Issledovatel'skogo Instituta gidrotekhniki Im. B.E. Vedeneeva}, 2000, vol. 236, pp. 121--133.

\item
 Danilova K.N., Liapidevskii V.Yu. {\it Uedinennye volny v dvukhsloinoi melkoi vode}, {\it Vestn. Novosib. Gos. Univ. Ser. Mat., Mekh., Inform.}, 2014, vol. 14, no. 4, pp. 22--31.

\item
 Khakimzyanov G.S., Gusev O.I., Beizel S.A., Chubarov L.B., Shokina N.Y. Simulation of tsunami waves generated by submarine landslides in the Black Sea, {\it Russian Journal of Numerical Analysis and Mathematical Modelling}, 2015, vol. 30, no .4, pp. 227--237.

\item
 Liapidevskii V.Yu. Shallow-water equations with dispersion. Hyperbolic model, {\it Journal of Applied Mechanics and Technical Physics}, 1998, vol. 39, no. 2, pp. 194--199.

\item
 Zeitunyan R.Kh. {\it Nelineinye dlinnye volny na poverkhnosti vody i solitony}, {\it Uspekhi Fizicheskikh Nauk}, 1995, vol. 165, no. 12, pp. 1403--1456.

\item
 Gusev O.I., Shokina N.Yu., Kutergin V.A., Khakimzyanov G.S. Numerical modelling of surface waves generated by underwater landslide in a reservoir, {\it Vychisl. Tecknol.}, 2013, vol. 18, no. 5, pp. 74--90 (in Russian).

\item
 Horritt M.S., Di Baldassarre G., Bates P.D., Brath A. Comparing the performance of a 2-D finite element and a 2-D finite volume model of floodplain inundation using airborne SAR imagery, {\it Hydrological Processes}, 2007, vol. 21, no. 20, pp. 2745--2759.

\item
 Bolgov M.V., Krasnozhon G.F., Shatalova K.Y. Computer hydrodynamic model of the Lower Volga, {\it Water resources}, 2014, vol. 41, no. 1, pp. 19--31.

\item
 Pisarev A.V., Khrapov S.S, Agafonnikova E.O., Khoperskov A.V. Numerical model of shallow water dynamics in the channel of the Volga: estimation of roughness, {\it Vestn. Udmurt. Univ. Mat. Mekh. Komp'yut. Nauki}, 2013, no. 1, pp. 114--130 (in Russian).

\item
 Caviedes-Voulli\'{e}me D., Morales-Hernandez M., Lopez-Marijuan I., Garc\'{\i}a-Navarro P. Reconstruction of 2D river beds by appropriate interpolation of 1D cross-sectional information for flood simulation, {\it Environmental Modelling \& Software}, 2014, vol. 61, pp. 206--228.

\item
 Costabile P., Costanzo C., Macchione F. A storm event watershed model for surface runoff based on 2D fully dynamic wave equations, {\it Hydrological Processes}, 2013, vol. 27, no. 4, pp. 554--569.

\item
 Juez C., Caviedes-Voulli\'{e}me D., Murillo J., Garc\'{\i}a-Navarro P. 2D dry granular free-surface transient flow over complex topography with obstacles. Part II: Numerical predictions of fluid structures and benchmarking, {\it Computers \& Geosciences}, 2014, vol. 73, pp. 142--163.

\item
 Marchuk A.G., Moshkalev P.S. Numeric simulation of the tsunami runup process on the shore with arbitrary profile, {\it Vestn. Novosib. Gos. Univ. Ser. Inform. Tecknol.}, 2014, vol. 12, no. 2, pp. 55--63 (in Russian).

\item
 Vacondio R., Rogers B.D., Stansby P.K. Smoothed Particle Hydrodynamics: Approximate zero-consistent 2-D boundary conditions and still shallow-water tests, {\it International Journal for Numerical Methods in Fluids}, 2012, vol. 69, no. 1, pp. 226--253.

\item
 Bautin S.P., Deryabin S.L., Sommer A.F., Khakimzyanov G.S., Shokina N.Yu. Use of analytic solutions in the statement of difference boundary conditions on a movable shoreline, {\it Russian Journal of Numerical Analysis and Mathematical Modelling}, 2011, vol. 26, no. 4, pp. 353--377.

\item
 Burguete J., Garc\'{\i}a-Navarro P. Implicit schemes with large time step for non-linear equations: application to river flow hydraulics, {\it International Journal for Numerical Methods in Fluids}, 2004, vol. 46, no. 6, pp. 607--636.

\item
 Burguete J., Garc\'{\i}a-Navarro P., Murillo J. Numerical boundary conditions for globally mass conservative methods to solve the shallow-water equations and applied to river flow, {\it International journal for numerical methods in fluids}, 2006, vol. 51, no. 6, pp. 585--615.

\item
 Burguete J., Garc\'{\i}a-Navarro P., Aliod R. Numerical simulation of runoff from extreme rainfall events in a mountain water catchment, {\it Natural Hazards and Earth System Science}, 2002, vol. 2, no. 1/2, pp. 109--117.

\item
 Westoby M.J., Glasser N.F., Brasington J., Hambrey M.J., Quincey D.J., Reynolds J.M. Modelling outburst floods from moraine-dammed glacial lakes, {\it Earth-Science Reviews}, 2014, vol. 134, pp. 137--159.

\item
 Singh J., Altinakar M.S., Ding Y. Numerical Modeling of Rainfall-Generated Overland Flow Using Nonlinear Shallow-Water Equations, {\it Journal of Hydrologic Engineering}, 2014, vol. 20, no. 8, p. 04014089.

\item
 Skiba Y.N. Finite-difference mass and total energy conserving schemes for shallow-water equations, {\it Rus. Meteorology and Hydrology}, 1995, vol. 2, pp. 55--65.

\item
 Ilgamov M.A., Gilmanov A.N. {\it Neotrazhayuschie usloviya na granitsakh raschetnoi oblasti} (Non-Reflecting Boundary Conditions for Computational Domains), Moscow: Fizmatlit, 2003, 240 p.

\item
 Zokagoa J.M., Soulaimani A. Modeling of wetting-drying transitions in free surface flows over complex topographies, {\it Computer Methods in Applied Mechanics and Engineering}, 2010, vol. 199, no. 33, pp. 2281--2304.

\item
 Liang Q., Borthwick A.G.L. Adaptive quadtree simulation of shallow flows with wet-dry fronts over complex topography, {\it Computers \& Fluids}, 2009, vol. 38, no. 2, pp. 221--234.

 \item
 Kopysov S.P., Tonkov L.E., Chernova A.A., Sarmakeeva A.S. Modeling of the incompressible liquid flow interaction with barriers using VOF and SPH methods, {\it Vestn. Udmurt. Univ. Mat. Mekh. Komp'yut. Nauki}, 2015, vol. 25, no. 3, pp. 405--420 (in Russian).

\item
 Vater S., Beisiegel N., Behrens J. A Limiter-Based Well-Balanced Discontinuous Galerkin Method for Shallow-Water Flows with Wetting and Drying: One-Dimensional Case, {\it Advances in Water Resources}, 2015, vol. 85, pp. 1--13.

\item
 Ostapenko V.V. Modified shallow water equations which admit the propagation of discontinuous waves over a dry bed, {\it Journal of Applied Mechanics and Technical Physics}, 2007, vol. 48, no. 6, pp. 795--812.

\item
 Khrapov S.S., Khoperskov A.V., Kuz’min N.M., Pisarev A.V., Kobelev I.A. A numerical scheme for simulating the dynamics of surface water on the basis of the combined SPH-TVD approach , {\it Vychisl. Metody Program.}, 2011, vol. 12, no. 2, pp. 282--297 (in Russian).

 \item
 Pisarev A.V., Khrapov S.S., Voronin A.A., Dyakonova T.A., Tsyrkova E.A. The role of infiltration and evaporation in the flooding dynamics of the volga-akhtuba floodplain, {\it Vеstnik Volgogradskogo Gosudarstvеnnogo Univеrsitеta. Sеriya 1, Matеmatika. Fizika}, 2012, no. 1 (16), pp. 36--41 (in Russian).

 \item
 Dyakonova T.A., Pisarev A.V., Khoperskov A.V., Khrapov S.S. Mathematical model of surface water dynamics, {\it Vеstnik Volgogradskogo Gosudarstvеnnogo Univеrsitеta. Sеriya 1, Matеmatika. Fizika}, 2014, no. 1 (20), pp. 35--44 (in Russian).

\item
 Khrapov S., Pisarev A., Kobelev I., Zhumaliev A., Agafonnikova E., Losev A., Khoperskov A. The numerical simulation of shallow water: estimation of the roughness coefficient on the flood stage, {\it Advances in Mechanical Engineering}, 2013, vol. 5, p. 787016.

  \item
 Shushkevich T.S., Kuz'min N.M., Butenko M.A. The three-dimensional parallel numerical code on the base of mixed lagrange-eulerian approach, {\it Vеstnik Volgogradskogo Gosudarstvеnnogo Univеrsitеta. Sеriya 1, Matеmatika. Fizika}, 2015, no. 4 (29), pp. 24--34 (in Russian).

\item
 Yee H.C., Beam R.M., Warming R.F. Boundary approximations for implicit schemes for one-dimensional inviscid equations of gasdynamics, {\it AIAA Journal}, 1982, vol. 20, no. 9, pp. 1203--1211.

\item
 Jin M., Fread D.L. Dynamic flood routing with explicit and implicit numerical solution schemes, {\it Journal of Hydraulic Engineering}, 1997, vol. 123, no. 3, pp. 166--173.

\item
 Cozzolino L., Della Morte R., Cimorelli L., Covelli C, Pianese D. A broad-crested weir boundary condition in finite volume shallow-water numerical models, {\it Procedia Engineering}, 2014, vol. 70, pp. 353--362.

\end{enumerate}

{ 

\renewcommand{\refname}{{\rm\centerline{СПИСОК ЛИТЕРАТУРЫ}}}

}

\end{document}